\documentclass{PoS}

              \usepackage{subcaption}                    
\usepackage{latexsym} 	
\usepackage{amsmath}  	
\usepackage{amsbsy}
\usepackage{epsfig}     
\usepackage{cite}

\usepackage{epsfig}
\usepackage{graphicx}
\usepackage{dcolumn}
\usepackage{bm}

\newcommand{\be}{\begin{equation}}
\newcommand{\ee}{\end{equation}}
\newcommand{\bea}{\begin{eqnarray}}
\newcommand{\eea}{\end{eqnarray}}

\title{Chiral phase transition of $N_f$=2+1 and 3 QCD at vanishing baryon chemical potential}

\ShortTitle{Chiral phase transition of $N_f$=2+1 and 3 QCD at vanishing baryon chemical potential}

\author{\speaker{Heng-Tong Ding} and Prasad Hegde for the Bielefeld-BNL-CCNU collaboration\thanks{ 
The numerical simulations were carried out on clusters of
the USQCD Collaboration in Jefferson Lab, and on TianHe I and II supercomputers at National Supercomputing Centers in Tianjin and Guangzhou.
PH is partly supported by Research Fund No. 11450110399 for International Young Scientists from National Natural Science Foundation of China. }\\\
        Key Laboratory of Quark \& Lepton Physics (MOE) and Institute of Particle Physics, \\
        Central China Normal University, Wuhan 430079, China\\
        E-mail: \email{hengtong.ding, phegde@mail.ccnu.edu.cn}}

\abstract{We present updated results on chiral phase structure in (2+1)-flavor ($N_f$=2+1) and 3-flavor ($N_f=3$) QCD 
based on the simulations using Highly Improved 
Staggered Quarks on lattices with temporal extent $N_\tau$ =6 at vanishing baryon chemical potential. In $N_f$=2+1 QCD we 
have performed simulations with a strange quark fixed to its physical value and two degenerate light quarks 
whose values are adjusted to have 5 values of Goldstone pion masses in the region of 160 - 80 MeV in the continuum limit.  
The universal scaling behavior of chiral condensates as well as chiral susceptibilities is discussed and 
the tri-critical point is suggested to be located below the physical point, i.e. at smaller than physical strange quark mass. 
In $N_f$=3 QCD simulations with 6 different masses of 3 degenerate quarks corresponding to the Goldstone pion masses in the region of 230 - 80 MeV 
have also been performed. Our results suggest that the QCD transition with these values of quark masses is of crossover type and
an upper bound  of the critical pion mass where the first order phase transition starts is estimated to be about 50 MeV.
}

\FullConference{The 33rd International Symposium on Lattice Field Theory\\
		14 -18 July 2015\\
		Kobe International Conference Center, Kobe, Japan*}

\begin{document}

\maketitle

\section{Introduction}

\begin{figure}[htp]
\begin{center}
\includegraphics[width=0.29\textwidth]{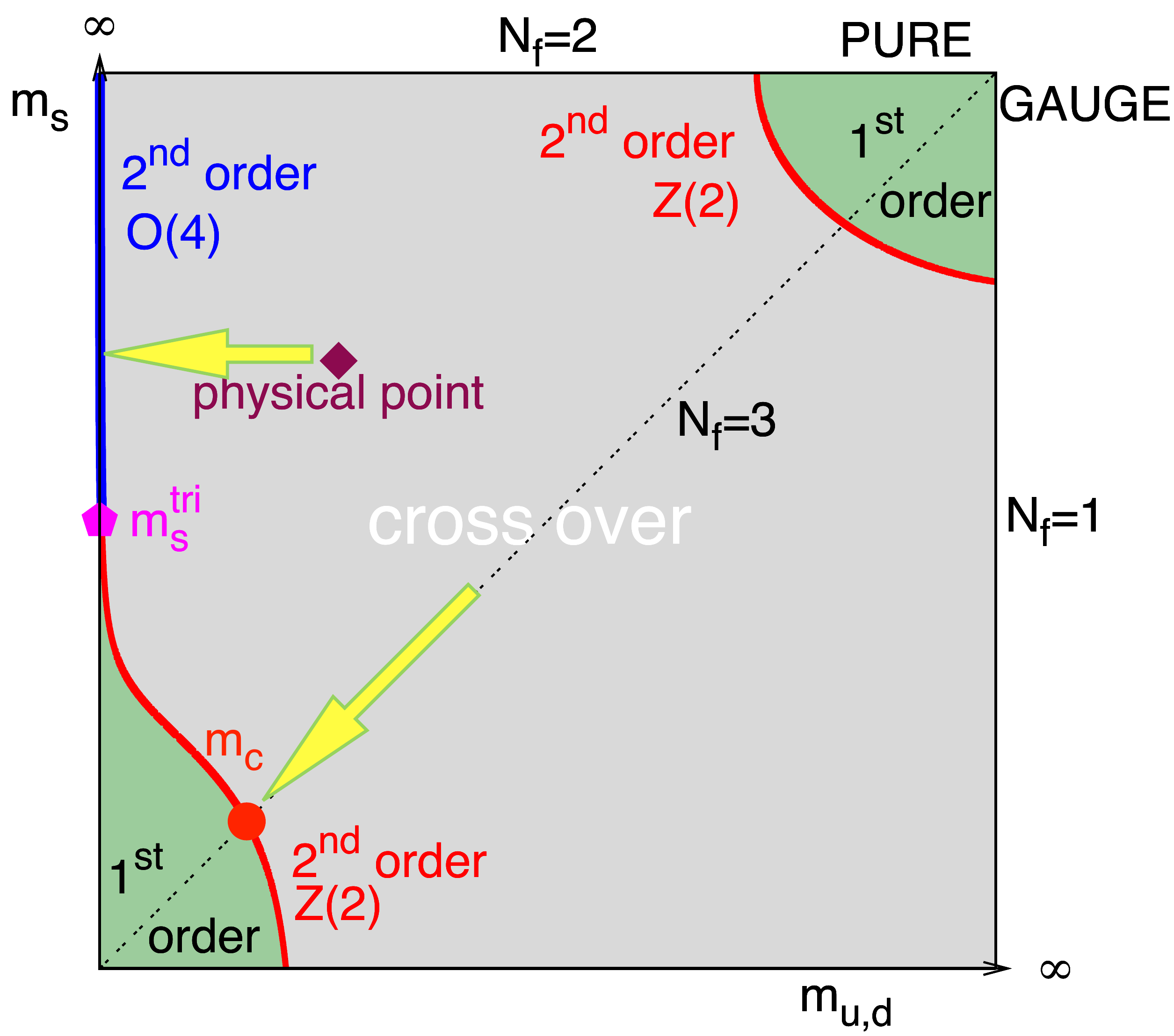}~~~
\end{center}
\caption{Schematic QCD phase structure with different values of quark 
masses ($m_{u,d}$, $m_{s}$) at zero baryon number density.}
\label{fig:sketch}
\end{figure}

The understanding of the QCD phase structure is 
one of the basic goals of lattice QCD computations at non-zero temperature.
It has been noted by Pisarski and Wilczek quite a long time ago that the QCD phase structure 
may depend on the number of light quark degrees of freedom~\cite{Pisarski:1983ms}.
The QCD phase structure in the quark mass plane is summarized in the so-called Columbia plot~\cite{Brown:1990ev}, i.e.
Fig.~\ref{fig:sketch}. The detailed description of the plot can be seen from e.g. Ref.~\cite{Ding:2015ona}.

In this proceedings we report on the updated studies of chiral phase structure
of (2+1)-flavor ($N_f$=2+1) and 3-flavor ($N_f$=3) QCD from the lattice QCD simulations with values of quark masses decreasing
towards the chiral limit along the yellow horizontal and diagonal arrows shown in Fig.~\ref{fig:sketch}, respectively.
The simulations have been performed using the Highly Improved Staggered Quarks on $N_\tau$=6 lattices.
We will mainly  discuss the relative position of the tri-critical point $m_s^{tri}$ and the physical point $m_s^{phy}$, and the
estimate of the critical quark mass $m_c$ as sketched in Fig.~\ref{fig:sketch}.
Previous studies were reported in Ref.~\cite{Ding:2013nha}.

\section{Lattice setup}

We performed simulations of $N_f$=2+1 and 3 QCD with lattice
spacings corresponding to temporal lattice size $N_\tau=6$ using
the Highly Improved Staggered fermions (HISQ).
In the case of $N_f=2+1$ QCD the strange quark mass is chosen to be fixed to its physical value ($m^{phy}_s$) 
and five values of light quark masses ($m_l$) that are varied in the range of 1/20$\gtrsim m_l/m^{phy}_s \gtrsim$1/80. 
These values of quark masses correspond to the lightest pseudo Goldstone pion masses of about 160, 140, 110, 90 and 80 MeV
in the continuum limit. To ensure $m_\pi L \gtrsim 4$ simulations are performed 
with $N_s$=24, 32 and 40 at light quark mass $m_l$=$m^{phy}_s/20$, $m^{phy}_s/40$ and $m^{phy}_s/60$, respectively. 
At our lowest quark mass corresponding to lightest Goldstone pion mass of about 80 MeV we also performed simulations at two different volumes of
$N_s=48$ and 32. Simulation parameters are listed in the left table in Table~\ref{tab:parameter}.

In the case of $N_f=3$ QCD we performed simulations with 3 degenerate quarks in which 
five different values of quark masses $am_q$  are varied from 0.0009375 to  0.0075
corresponding to pion masses in the region of $230 \gtrsim m_\pi \gtrsim 80$ MeV.
The aspect ratio $N_\sigma/N_\tau$ is generally 4, although other volumes i.e. $N_\sigma=16,12,10$ at $am_q=0.00375$
and $N_\sigma=16$ at $am_q=0.0009375$ have also been used. Simulation parameters are listed in the right table in Table~\ref{tab:parameter}.

\begin{table}[!htb]
    \begin{subtable}{.3\linewidth}
      \centering
        \tiny
   \begin{tabular}{|c|c|c|c|}
\hline
$N_\sigma^3\times N_\tau$ &   $m_l/m^{phy}_s$     & $m_{\pi}$ [MeV]      & average \# of conf. \\
\hline
$24^3\times$ 6           & 1/20          & 160                                        & 1200             \\
$12^3\times$ 6          & 1/27        & 140                                & 1200           \\
$16^3\times$ 6          & 1/27        & 140                                 & 1500            \\
$20^3\times$ 6          & 1/27        & 140                                & 1000             \\
$24^3\times$ 6          & 1/27        & 140                                   & 1000             \\
$32^3\times$ 6          & 1/27        & 140                                 & 1300           \\
$32^3\times$ 6           & 1/40      & 110                                  &  1200           \\
$40^3\times$ 6          & 1/60        &  90                                     & 1000           \\
$32^3\times$ 6           & 1/80    & 80                                        & 1200           \\
$48^3\times$ 6           & 1/80    & 80                                       & 900           \\

\hline
        \end{tabular}
    \end{subtable}%
    \begin{subtable}{1.\linewidth}
      \centering
        \tiny
\begin{tabular}{|c|c|c|c|}
\hline
$N_\sigma^3\times N_\tau$&   $am_q$     & $m_{\pi}$ [MeV]     &   average \# of conf.  \\
\hline
$16^3\times$ 6           & 0.0075          & 230                                   & 1000             \\
$24^3\times$ 6          & 0.00375        & 160                                & 2000             \\
$16^3\times$ 6          & 0.00375        & 160                               & 2000             \\
$12^3\times$ 6          & 0.00375        & 160                               & 2000             \\
$10^3\times$ 6          & 0.00375        & 160                                  & 1500             \\
$24^3\times$ 6          & 0.0025        &  130                                     & 1300          \\
$24^3\times$ 6           & 0.001875      & 110                                    & 1000           \\
$24^3\times$ 6          & 0.00125        &  90                                       & 1000           \\
$24^3\times$ 6           & 0.0009375    & 80                                        & 1500        \\
$16^3\times$ 6        & 0.0009375    &  80                                  & 1500          \\
\hline
\end{tabular}
        
    \end{subtable} 
     \caption{Parameters of the numerical simulations for $N_f$=2+1 QCD (left) and $N_f=3$ QCD (right).}
\label{tab:parameter}
\end{table}

\section{Universality class near critical lines}

Close to the chiral limit the chiral order parameter ($M$) and its
susceptibility ($\chi_M$) can be described by the universal properties of the chiral
transition \cite{Ejiri:2009ac}, i.e. so called magnetic equation of state (MEOS),
\be
M(t,h)= h^{1/\delta} f_G(z)  + f_{reg}
\;, \qquad\mathrm{and}\qquad
\chi_M(t,h) = \frac{\partial M}{\partial H}=\frac{1}{h_0} \,h^{1/\delta -1}\,
f_\chi(z) + \chi_{reg}
\;.
\label{eq:MEOS}
\ee
where 
\be
t=\frac{1}{t_0} \,\frac{T-T_c}{T_c}, ~\qquad\mathrm{and}\,\;~h= \frac{H}{h_0}=\frac{1}{h_0} \frac{m_l-m_c}{m_s}.
\ee
$t$ and $h$ are reduced temperature and symmetry-breaking field, respectively.
They reflect the proximity of a system to a critical region.
Here $z=th^{-1/\beta\delta}$ is the scaling variable. The critical exponents $\beta$, $\delta$ and the scaling functions $f_G(z)$ and 
$f_\chi(z)$ uniquely characterize the universality class of the QCD chiral
phase transition. 

In the chiral limit of $N_f=2$ QCD the universality class 
is believed to be that of the 3-d O(4) spin model~\footnote{At the finite lattice cutoff for staggered fermions, as used in the current study, there is only one Goldstone boson in the
chiral limit, the relevant universality class is rather that of 3-d O(2) spin model.} and the critical quark mass $m_c=0$, while in the 3 degenerate flavor case
towards the chiral limit it would be that of a Z(2) spin model with the values of quark mass closer 
to a presumably nonzero critical quark mass $m_c$. The parameters $t_0$, $h_0$, $T_c$ are non-universal and specific for a
theory and $T_c$ here is a fundamental quantity, i.e. the phase transition temperature of QCD.
These non-universal parameters can be determined by studying the universal behavior of the order parameter and its corresponding
susceptibility described in Eq.~(\ref{eq:MEOS}).

\section{Chiral phase transition of $N_f$=2+1 QCD}

\begin{figure}[htp]
\begin{center}
\includegraphics[width=0.35\textwidth]{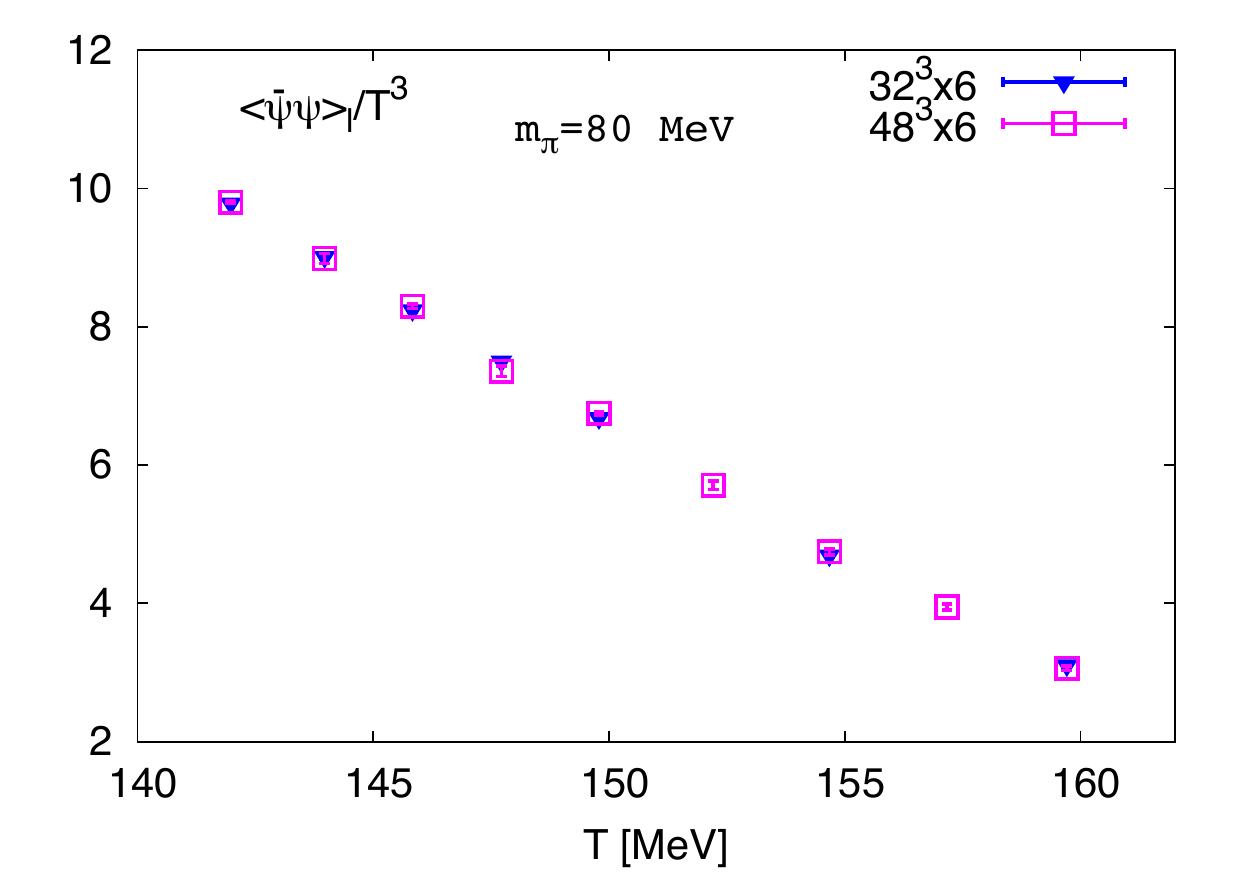}~~~
\includegraphics[width=0.35\textwidth]{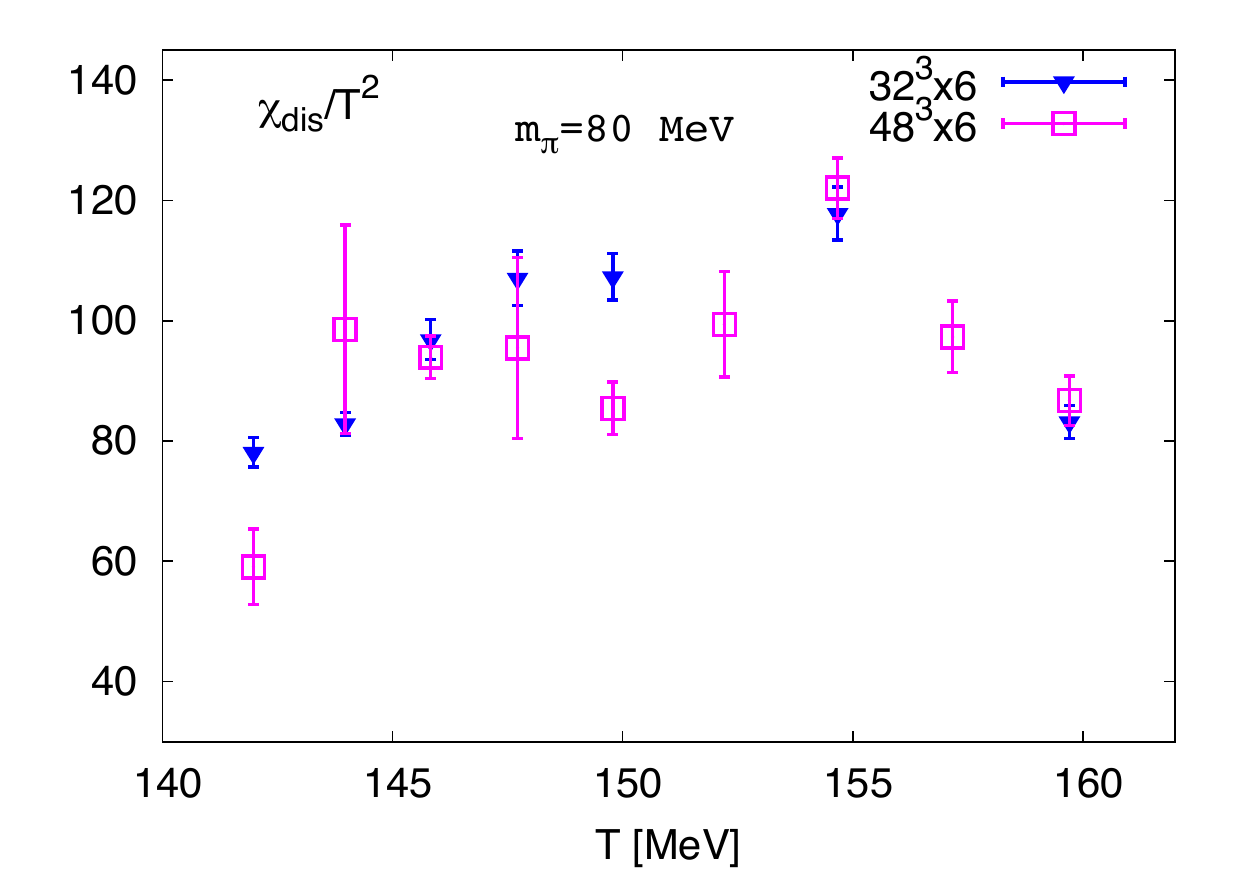}~
\end{center}
\caption{Volume dependences of light quark chiral condensates (left) and of disconnected chiral susceptibilities (right)
for the light quark mass $m_l=m_s^{phy}/80$ corresponding
to $m_\pi\simeq80~$MeV.}
\label{fig:Vdep_80MeV}
\end{figure}

In Fig.~\ref{fig:Vdep_80MeV} we show volume dependence of light quark chiral condensates $\langle\bar{\psi}\psi\rangle_l$ and disconnected chiral susceptibilities $\chi_{disc}$ at our lowest
light quark mass corresponding to $m_\pi=80 $ MeV in the continuum limit in the left and right plot, respectively. 
The chiral condensate has minor volume dependence and there is no sudden jump seen in the temperature dependence of the chiral condensate.
Together with minor volume dependence seen from the disconnected susceptibility it is clear that in the investigated
 quark mass window there does not exist a first order phase transition.

\begin{figure}[htp]
\begin{center}
\includegraphics[width=0.35\textwidth]{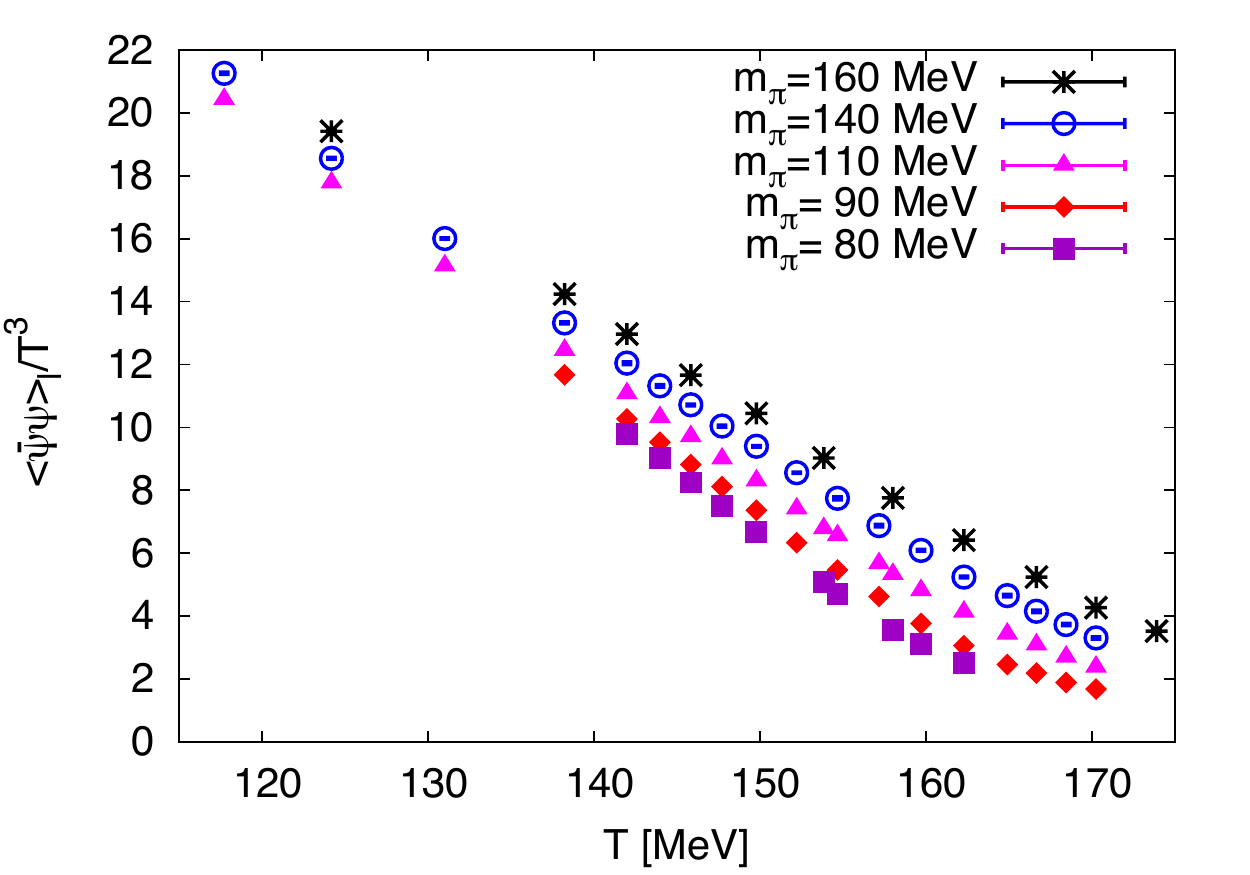}~~~
\includegraphics[width=0.35\textwidth]{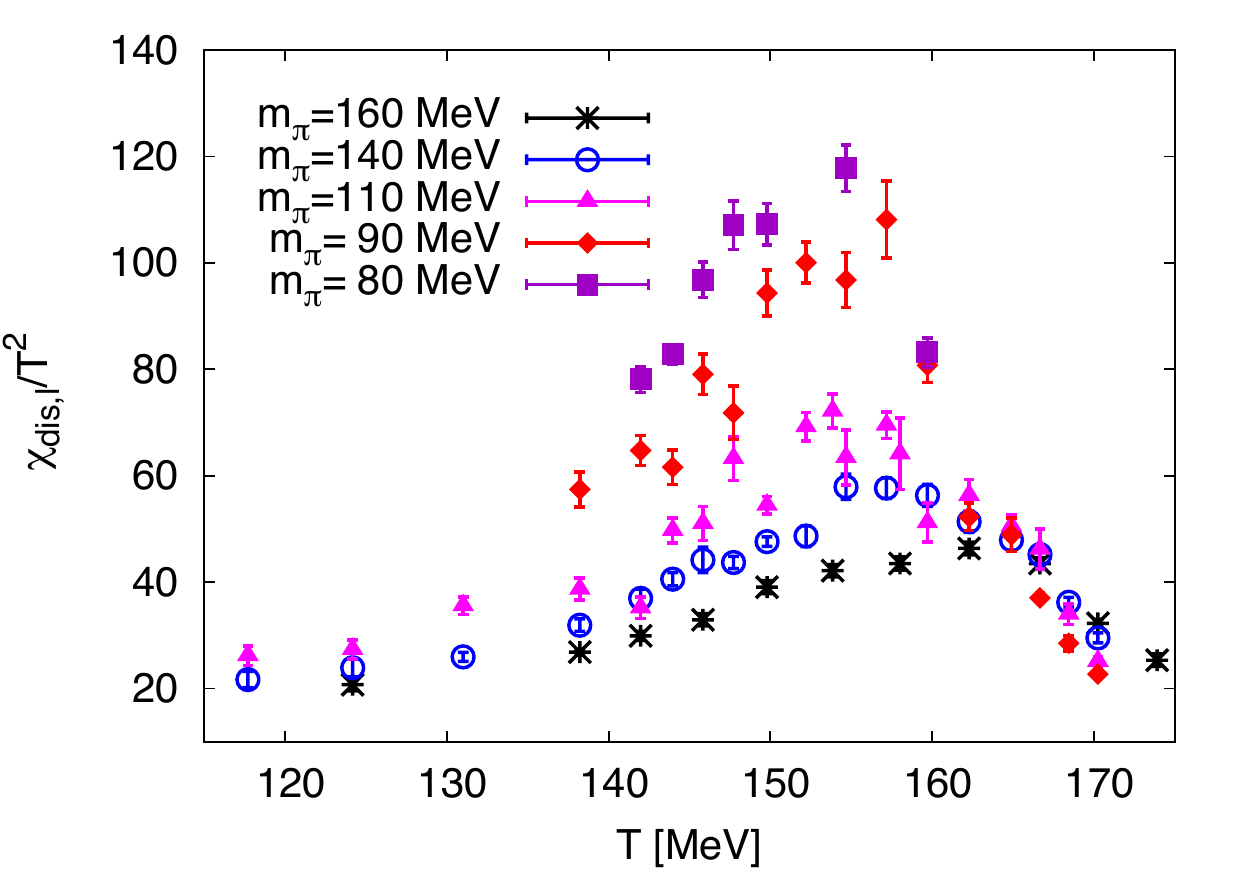}~
\end{center}
\caption{Quark mass dependences of light quark chiral condensates (left) and disconnected chiral susceptibilities (right).}
\label{fig:MdepNf21}
\end{figure}

\begin{figure}[htp]
\begin{center}
\includegraphics[width=0.35\textwidth]{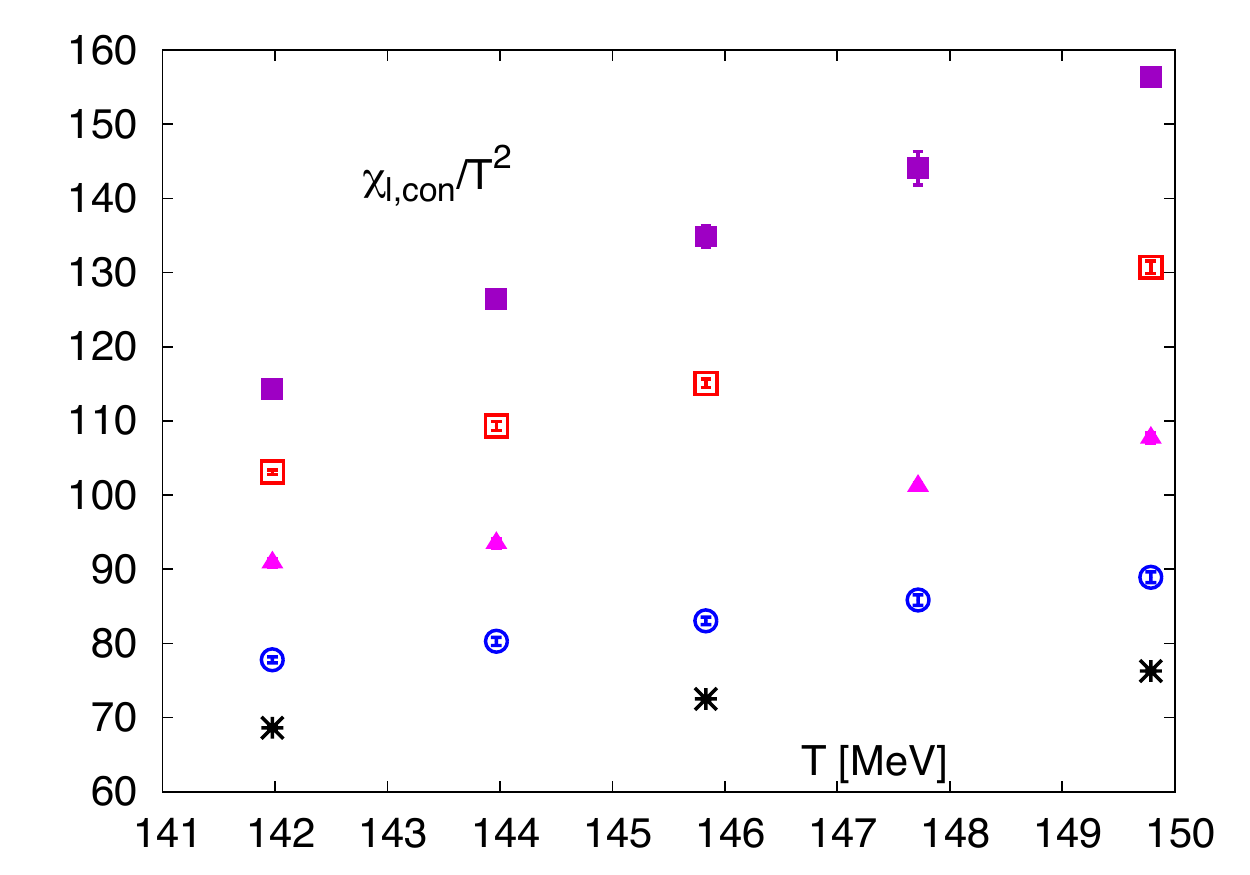}~~~
\includegraphics[width=0.35\textwidth]{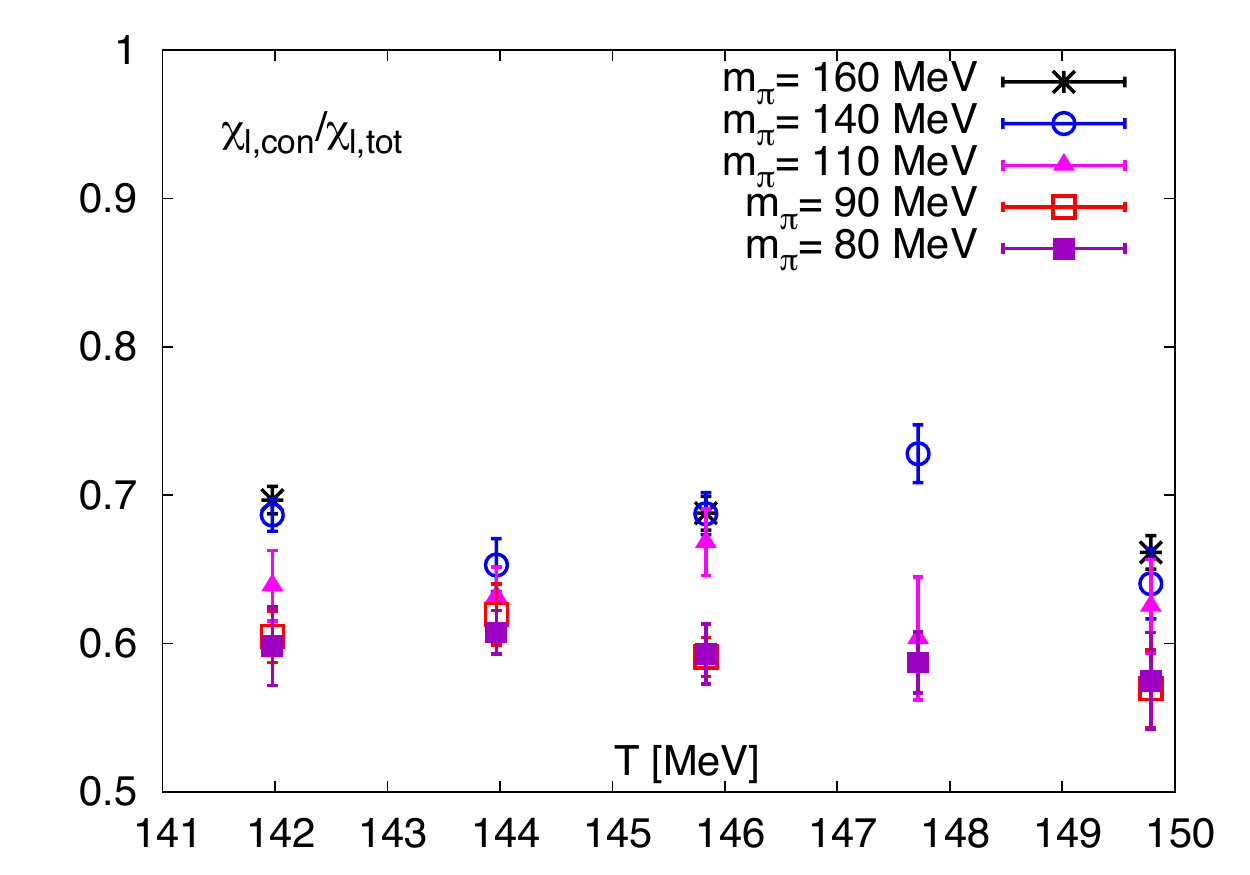}~
\\\end{center}
\caption{Left: Temperature dependence of connected susceptibility at various values of quark masses. 
Right: Ratio of the connected susceptibility
to the total susceptibility.}
\label{fig:chiconn}
\end{figure}

The mass dependence of chiral condensates and disconnected chiral susceptibilities
is shown in Fig.~\ref{fig:MdepNf21}. As expected the chiral condensate decreases
and the peak height of the disconnected susceptibility becomes larger at a smaller value of quark mass, since
the system approaches to a phase transition as the light quark masses go to zero. 
In the continuum 2-flavor QCD the connected quark susceptibility $\chi_{con}$ stays finite, however, for the staggered
quarks $\chi_{con}$ becomes divergent in the chiral limit. In Fig.~\ref{fig:chiconn} we show the temperature dependence
of connected susceptibility $\chi_{con} $ and the ratio $\chi_{con}/\chi_{total}$ at various values of quark masses.
It can be seen that at the temperature from about 142 MeV to about 150 MeV the connected susceptibility increases
with temperature and becomes larger at smaller values of quark masses. This trend is similar to the disconnected susceptibility.
The ratio of the connected susceptibility to the total susceptibility on the other hand is almost temperature independent in this
temperature region and it is also independent on the quark mass at the smaller quark masses. Besides that the connected susceptibility is larger than the
disconnected susceptibility at a certain set of temperature and quark mass. This may indicate
that singular contributions dominate in both two quantities but have different fractions in the total singular contributions.
When looking into the ratios the only thing left is the relative factor which is independent of quark mass and temperature.

\begin{figure}[htp]
\begin{center}
\includegraphics[width=0.24\textwidth]{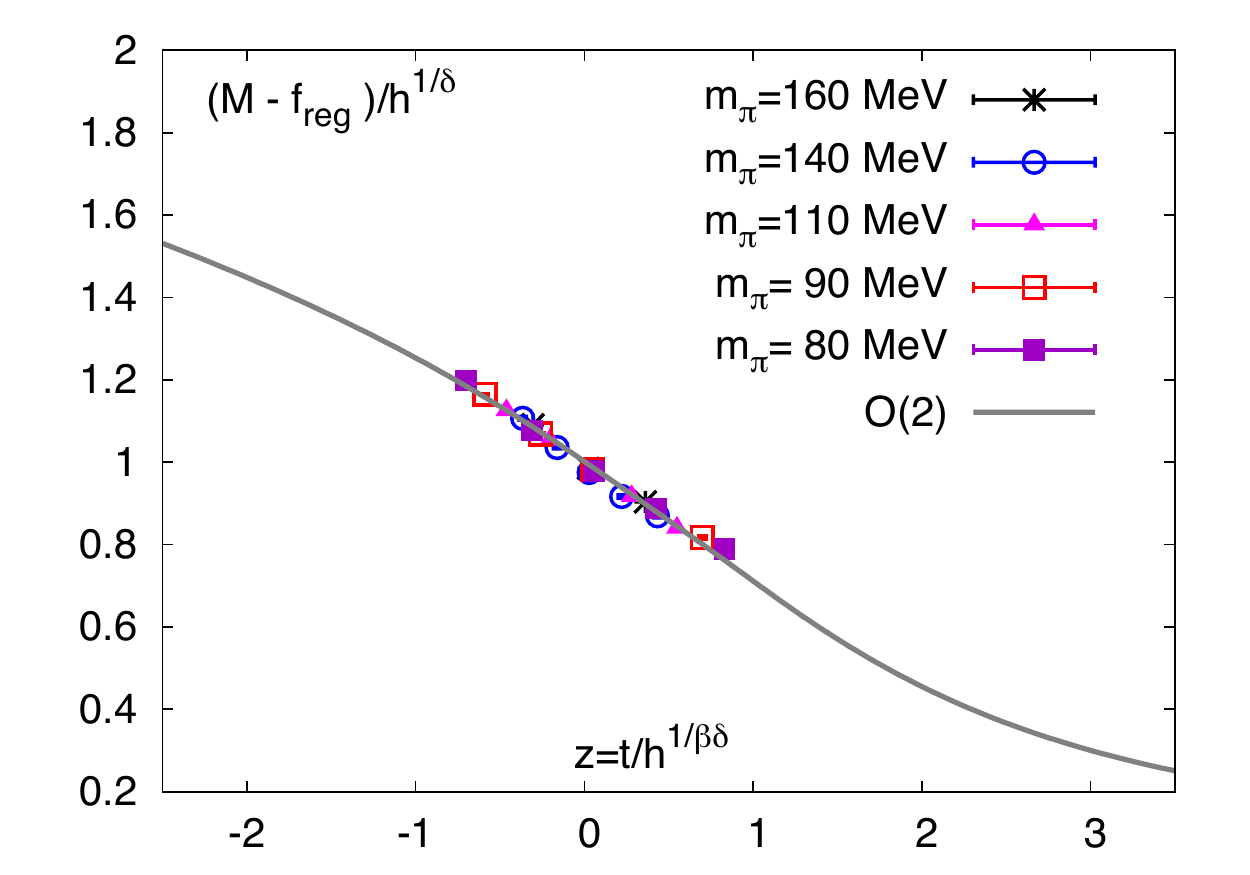}~
\includegraphics[width=0.24\textwidth]{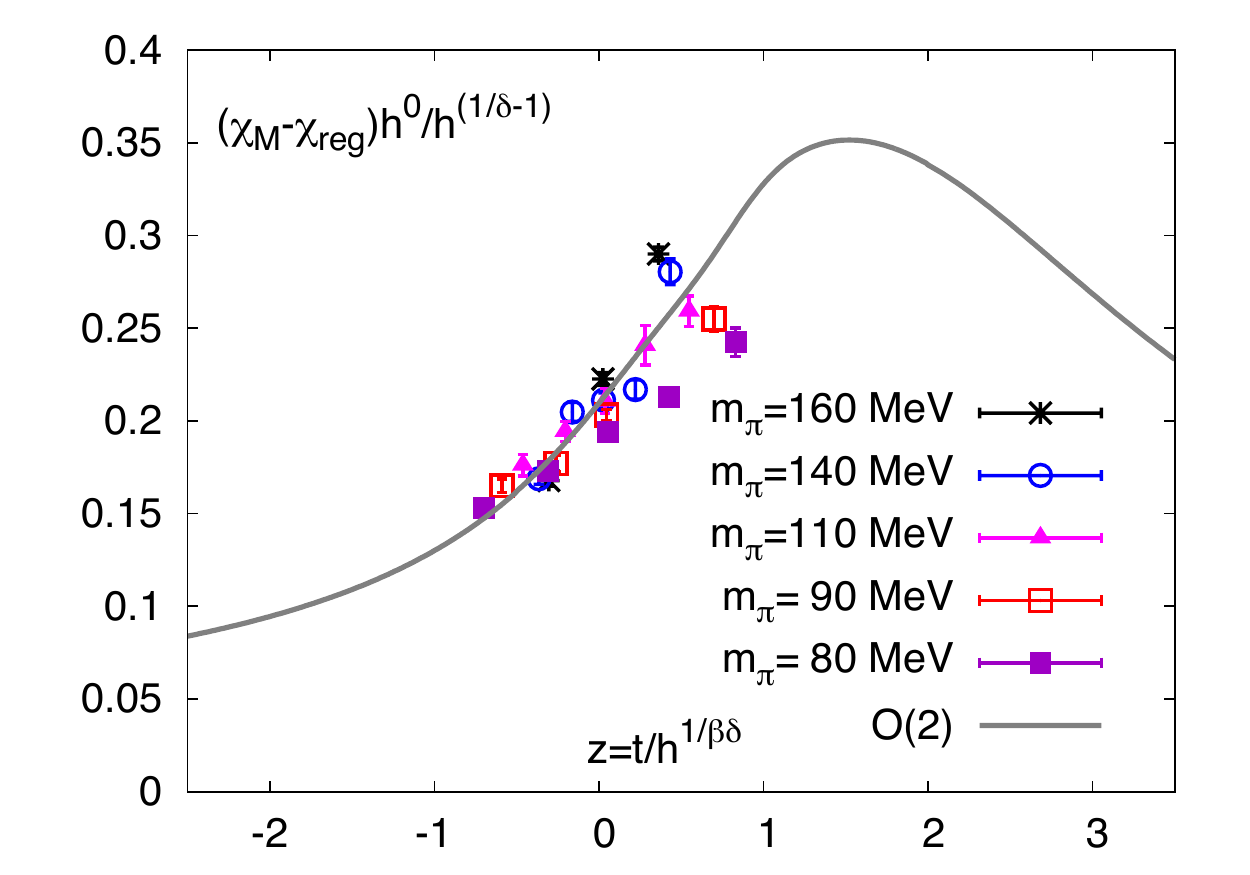}
\includegraphics[width=0.24\textwidth]{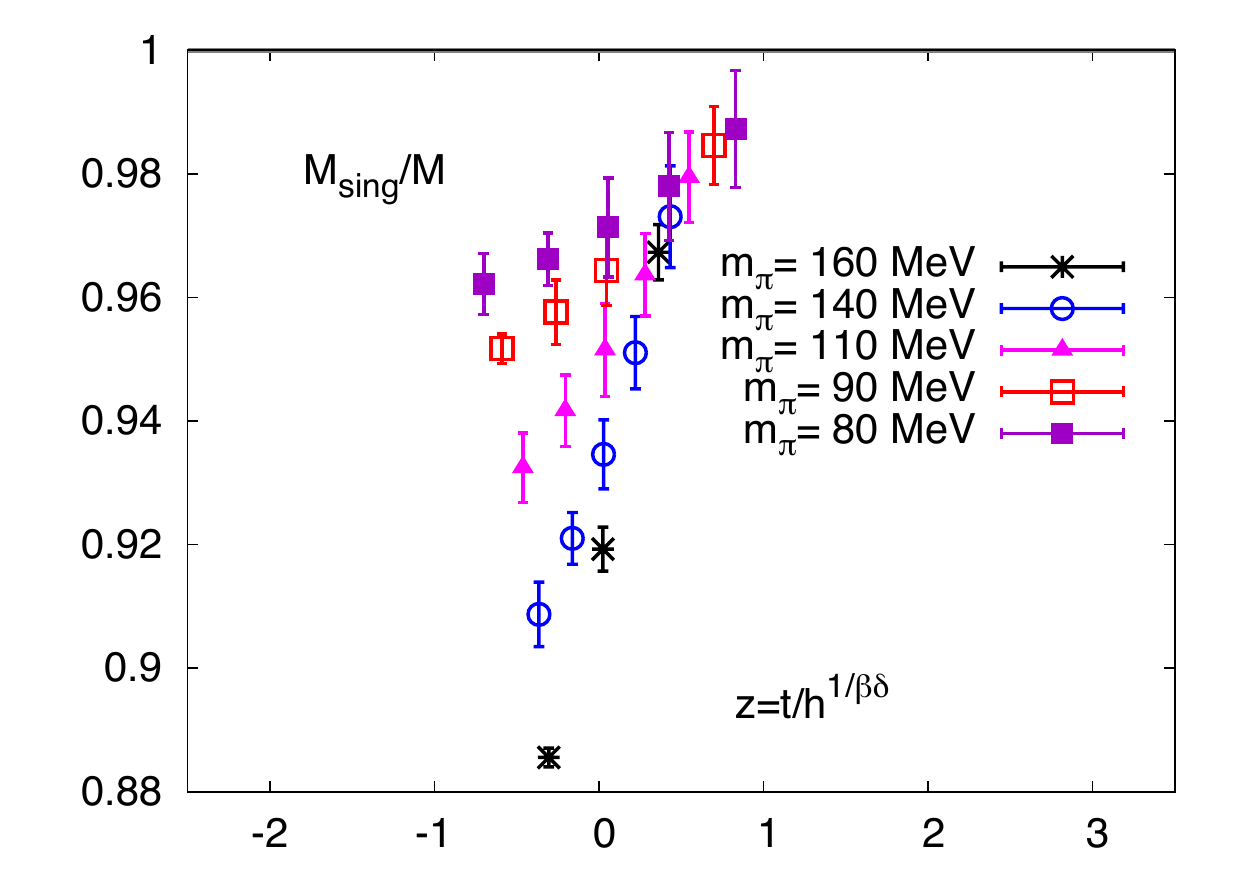}
\includegraphics[width=0.24\textwidth]{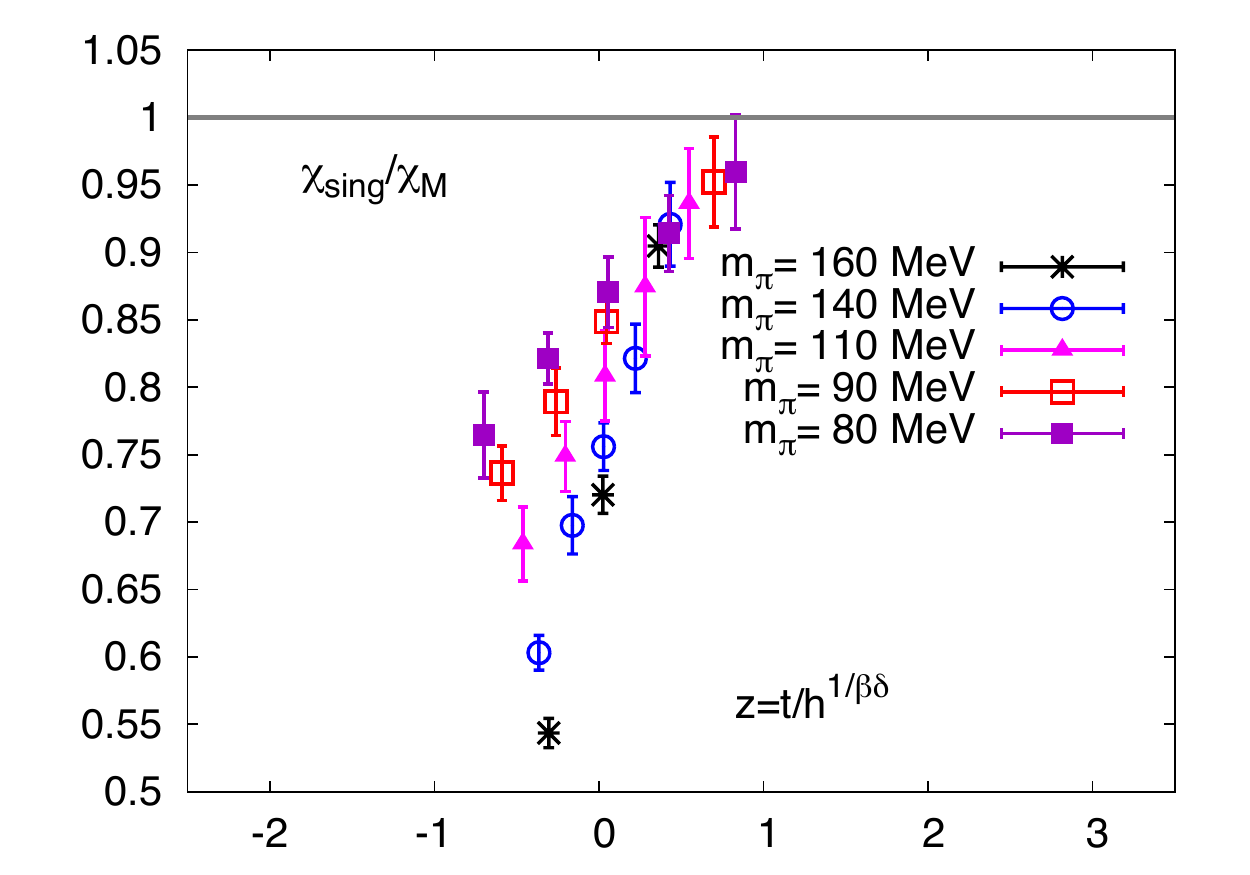}
\end{center}
\caption{Left half panel: O(2) scaling analysis for all quark masses obtained from the scaling fits to the order parameter 
$M=m_s\langle \bar\psi\psi\rangle_l N_\tau^4$  and $\chi_M=m_s^2\chi_{total}N_\tau^4$.
Right half panel: the resulting fraction of the contribution from the singular part in $M$ (left) and $\chi_M$ (right). }
\label{fig:MEOSO2}
\end{figure}
\begin{figure}[htp]
\begin{center}
\includegraphics[width=0.35\textwidth]{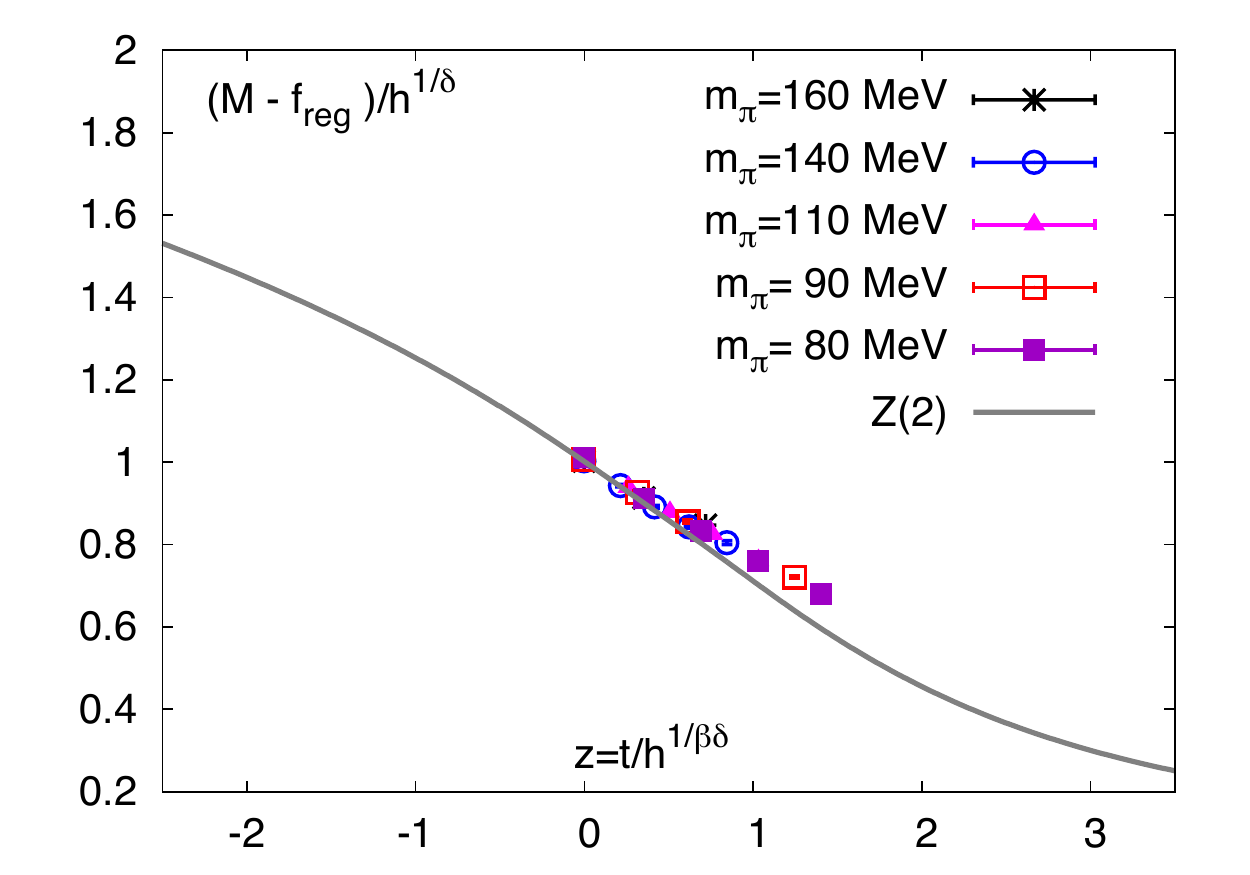}~
\includegraphics[width=0.35\textwidth]{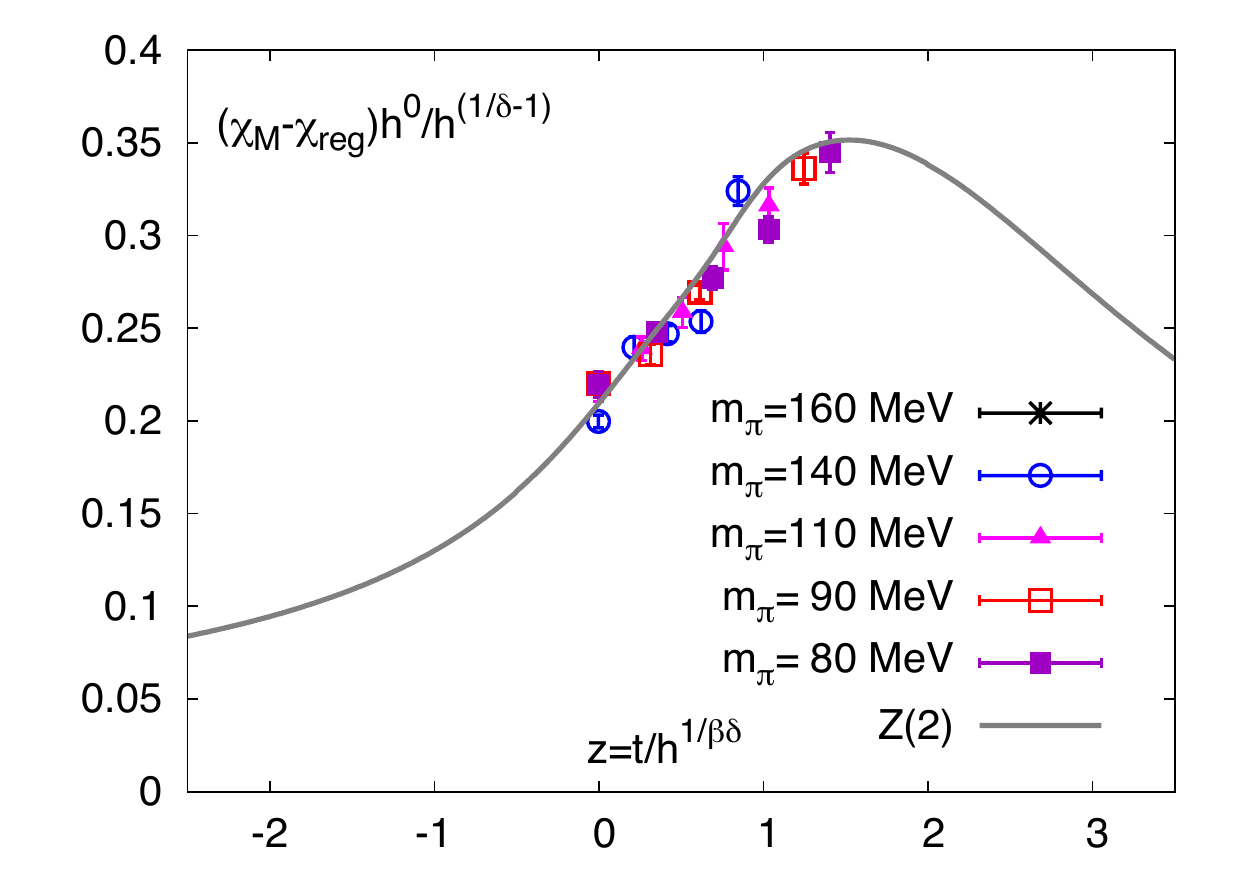}
\end{center}
\caption{Same as the plots in the left half panel in Fig. 5 but using the Z(2) universality class.}
\label{fig:MEOSZ2}
\end{figure}

To analyze the universal behavior of chiral phase transition in the chiral limit of two-flavor QCD we performed 
a joint scaling fit to chiral condensates and total chiral susceptibilities at all values of quark masses available using the MEOS 
according to Eq.~(\ref{eq:MEOS}). The scaling functions used in the fits belong to the O(2) universality class.
For the regular part we used $f_{reg}=\left(a_0 + a_1 \frac{T-T_c}{T_c} \right) \frac{m_l}{m_s}$ and $\chi_{reg}=a_0 + a_1 \frac{T-T_c}{T_c}$. 
The fitting results are shown in the two plots in the left hand side of Fig.~\ref{fig:MEOSO2}. We can see that the MEOS provides 
a good description to chiral condensates and the scaling curve passes through the data points of total chiral susceptibilities. 
The deviation in the total susceptibility at large quark mass from the scaling curve could be due to the fact that 
the regular contributions are larger there and one needs to include in the regular term the contributions from the
beyond-leading-order corrections of the quark mass, while the deviation from the data points at lowest quark mass
should be due to the poor statistics. We also show the fraction of the contribution from the singular part in the chiral condensate
and total chiral susceptibilities in the two plots in the right hand side of Fig.~\ref{fig:MEOSO2}. It is expected that the contribution from the singular part
dominates over that from the regular part more at the smaller quark mass as the system approaches to the critical regime. In fact, the ratio 
of $\chi_{sing}/\chi_{M}$ at $z\sim 0$, i.e. $T\sim T_c$ increases from $\sim 72$\% at $m_\pi=160$ MeV
to $\sim 88$\% at $m_\pi=80$ MeV.

\begin{table}[t]
\begin{center}
\small
\begin{tabular}{|c|c|c|c|c|c|c|}
\hline
Universality class &$m_c/m_{s}^{phy}$ &$T_c $ [MeV]&   $t_0$     & $h_0$    & $a_0$ & $a_1$ \\
\hline
O(2) &  N/A  &145.6(1)  & 3.34(4)e-03        &4.56(4)e-06                             & 10.1(3)      &   -384(10)  \\
Z(2) &-0.0077(2) & 141.9(2)    & 1.5(2)e-04        &3.54(8)e-06                                & 1.7(5)     &   --282(8) \\
\hline
\end{tabular}
\end{center}
\caption{Obtained non-universal parameters in $N_f=2+1$ QCD with O(2) and Z(2) universality classes.}
\label{tab:scaling}
\end{table}

In the previous O(2) scaling analysis we assumed that
the tri-critical point in the columbia plot is located below the physical point. It may 
happen that tri-critical point is above the physical point and the system passes through the
critical line belonging to Z(2) universality class towards the chiral limit of 2 light quarks.
We thus have performed a scaling fit to both chiral condensates and chiral susceptibilities
using also Z(2) scaling functions. The fit includes an additional fit parameter $m_c$. From the fit results
shown in Fig.~\ref{fig:MEOSZ2} a very small (even negative) value of $m_c$ is obtained. This certainly suggests
that the Z(2) scaling does not work for the $N_f$=2+1 QCD. The other non-universal parameters
are also listed in Table~\ref{tab:scaling}. 
Thus, this finding together 
with that from the O(2) scaling analysis suggest that the tri-critical point is indeed below the tri-critical point, i.e. $m_s^{tri}<m_s^{phy}$.

\section{Chiral phase structure of $N_f$=3 QCD}

\begin{figure}[htp]
\begin{center}
\includegraphics[width=0.24\textwidth]{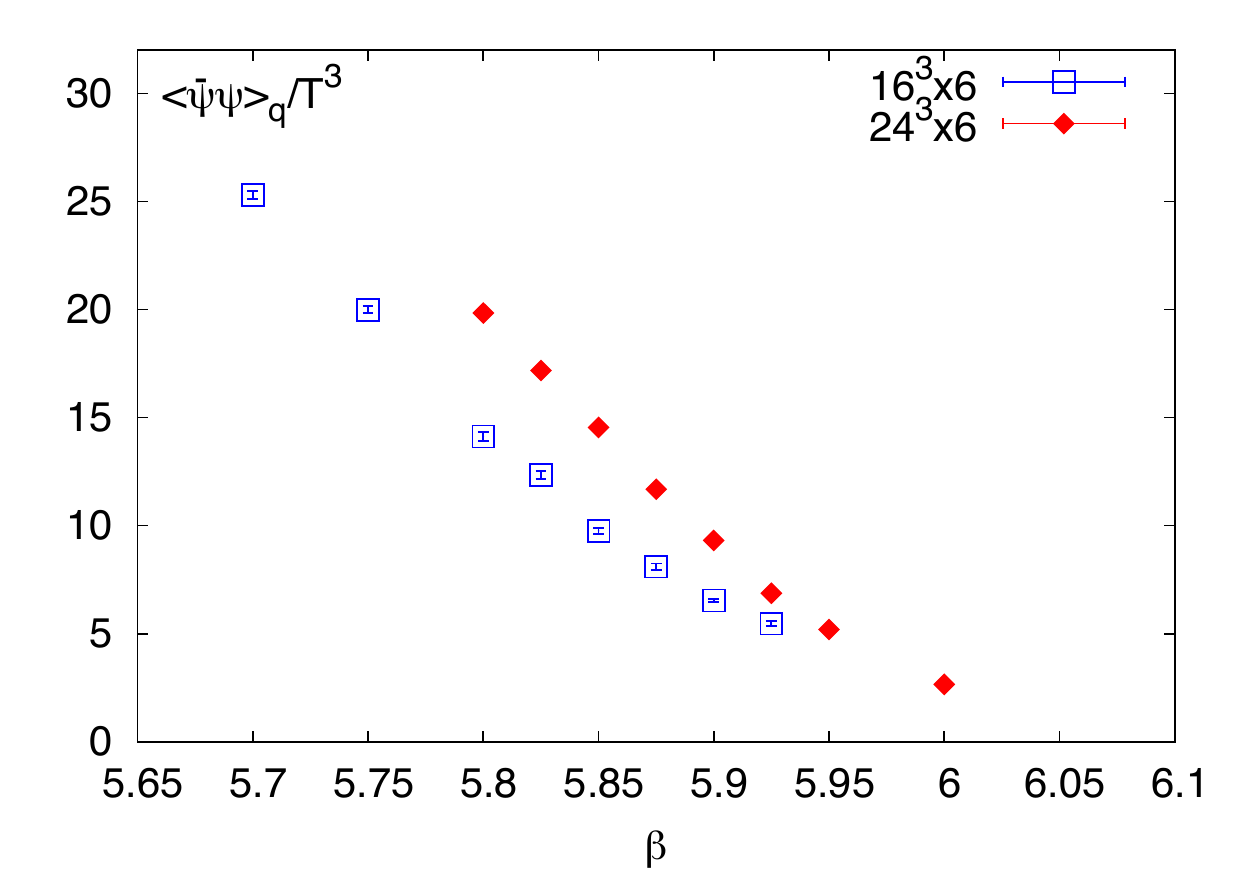}~\includegraphics[width=0.24\textwidth]{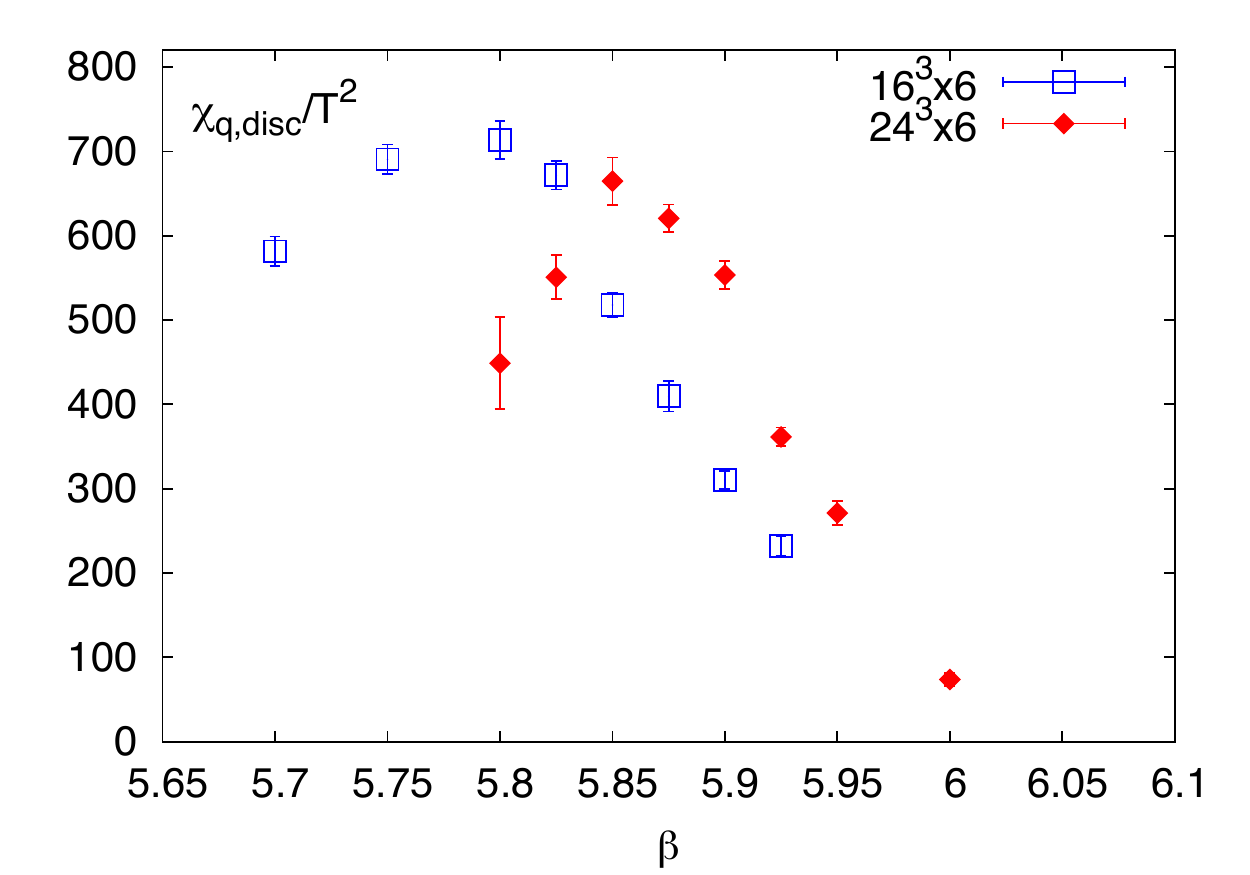}
\includegraphics[width=0.24\textwidth]{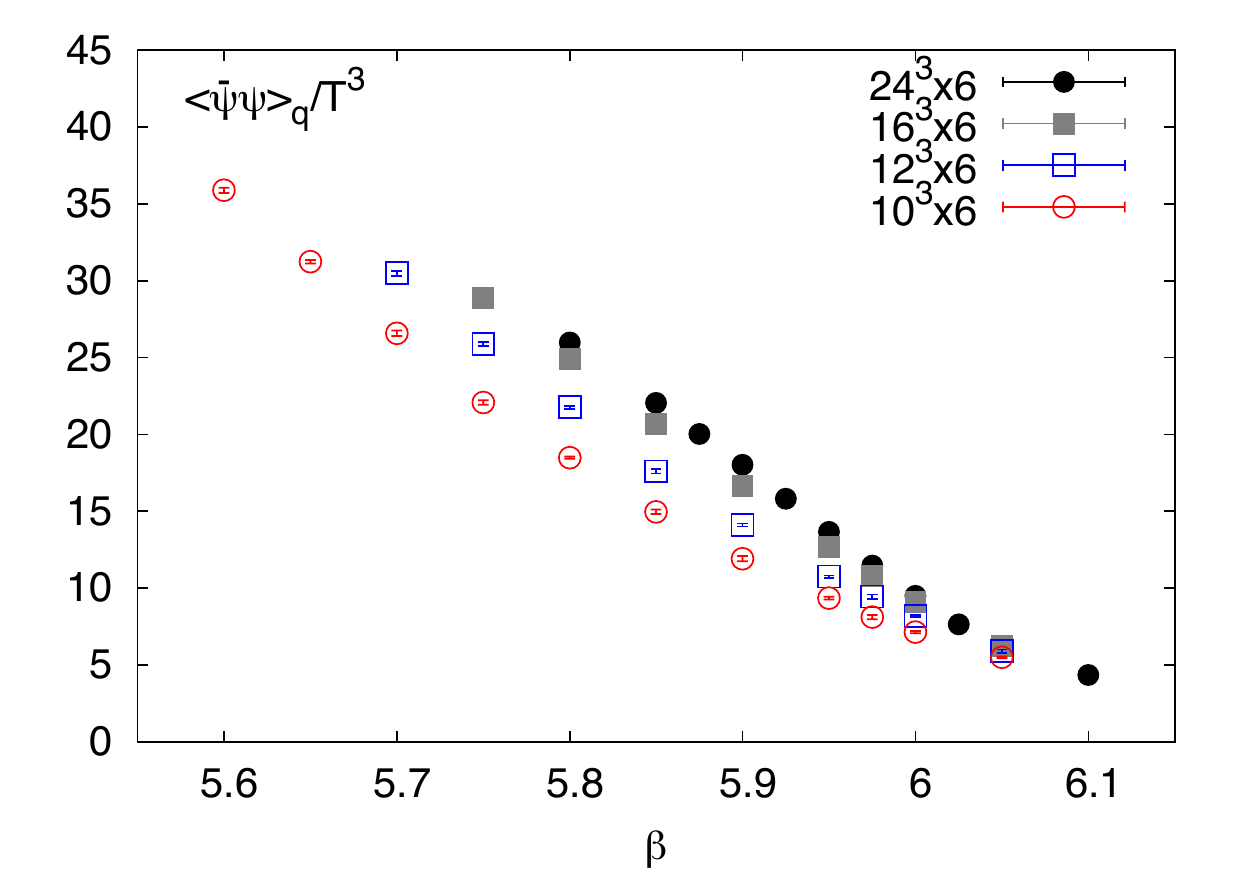}~\includegraphics[width=0.24\textwidth]{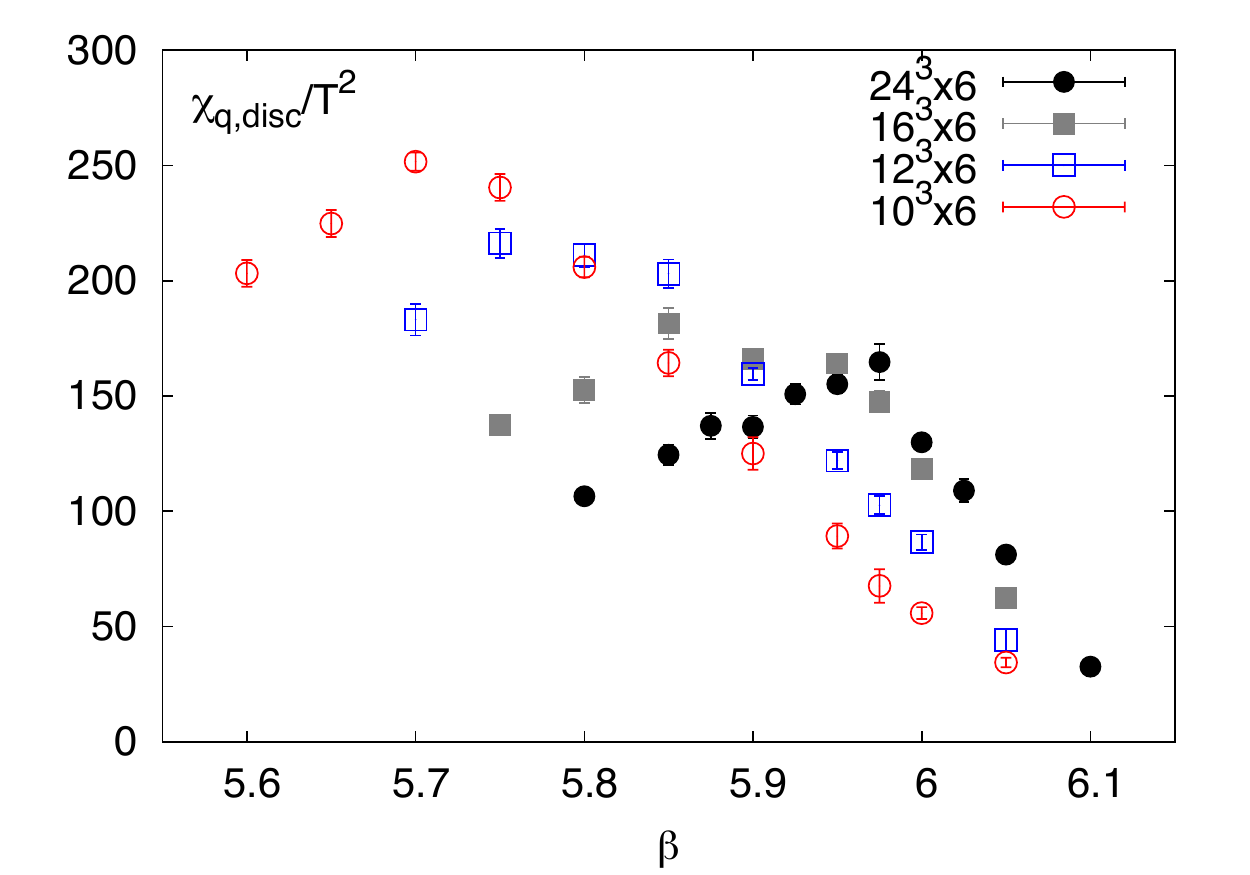}
\end{center}
\caption{Volume dependences of chiral condensates and disconnected chiral susceptibilities at $am_q$=0.0009375 (left half panel) and $am_q$=0.00375 (right half panel).}
\label{fig:Vdep3f}
\end{figure}

To investigate whether the system has reached the chiral first order phase transition region 
we have performed lattice QCD simulations with various mass values of 3 degenerate quarks.
We show the volume dependence of chiral condensates
and disconnected susceptibilities at lowest quark mass $am_q=0.0009375$ in the left half panel of Fig.~\ref{fig:Vdep3f}.
There is no discontinuity in the temperature dependence of chiral condensate and no linear increase of disconnected
chiral susceptibility in volume observed. This suggests that the system with investigated quark mass window, i.e. corresponding to $ 80 \lesssim m_\pi \lesssim 230$ MeV
 is not yet in the chiral first order phase transition region.

\begin{figure}[htp]
\begin{center}
\includegraphics[width=0.3\textwidth]{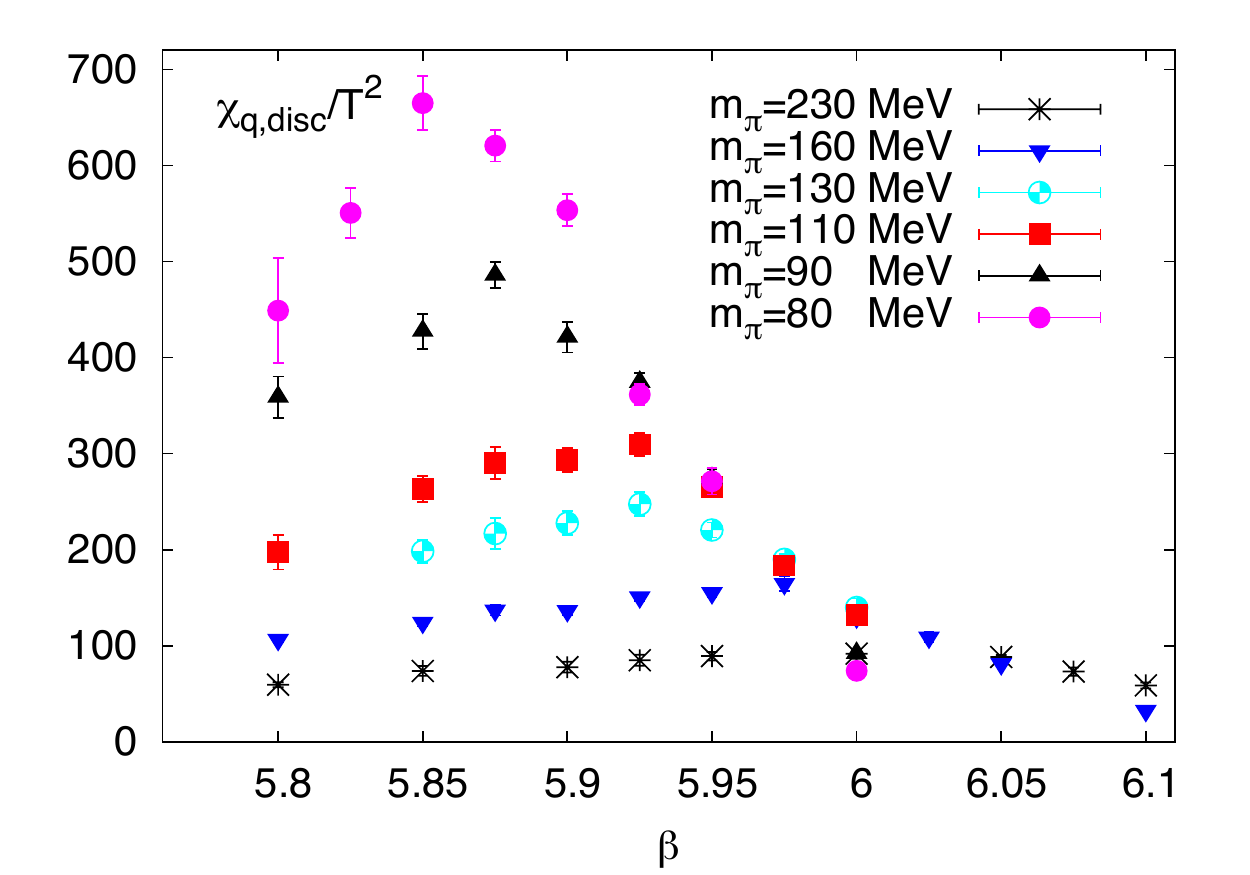}
\includegraphics[width=0.3\textwidth]{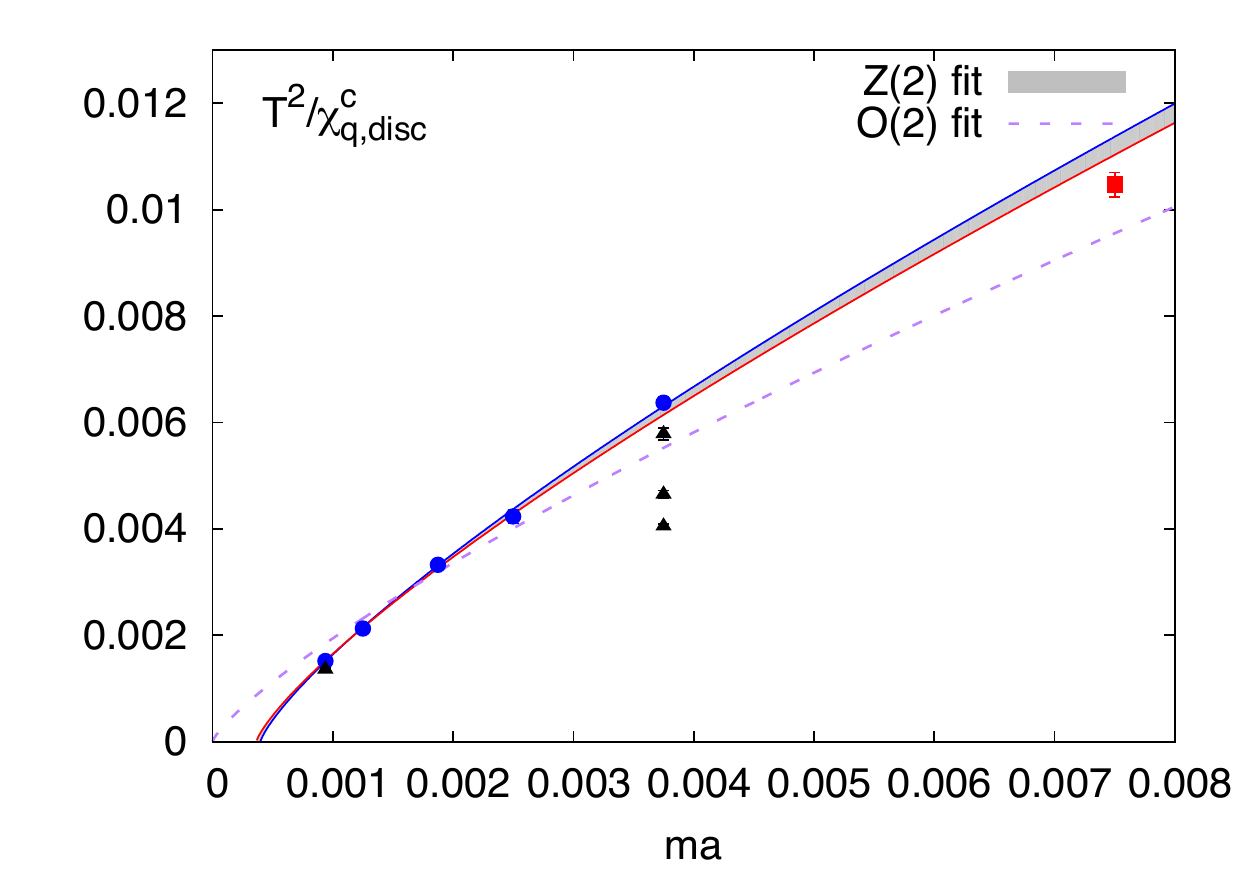}
\end{center}
\caption{Left: Disconnected susceptibilities from the largest available volume. Right: Estimate of the critical quark mass $m_c$ according to the Z(2) universality class. 
 }
\label{fig:Fit3f}
\end{figure}

To estimate to what value of quark mass the first order chiral phase transition region extends we
assume the system is in the criticality window belonging to the Z(2) universality class. We here use the chiral condensate 
as an approximation of the order parameter of the chiral phase transition of 3 degenerate favor QCD. Thus at the 
pseudo-critical temperature, i.e. peak location of the disconnected chiral susceptibility, the inverse disconnected susceptibility 
$T/\chi^{max}_{q,disc}$ is proportional to $(m-m_c)^{1-1/\delta}$. Thus an estimate of $m_c$ can be obtained by fitting a 2-parameter ansatz
to the inverse peak height of the disconnected susceptibility as can be read off from the left plot of Fig.~\ref{fig:Fit3f}. The resulting fit is shown in
the right plot of Fig.~\ref{fig:Fit3f} and the intercept of the fitting curve to the $x$ axis is the estimate of $am_c$. The grey band indicates the
uncertainties arising from the fit with and without taking the data point at largest quark mass into account. The data points obtained in the smaller volume
are shown as the black points which could make the estimate of $am_c$ smaller. Thus we obtained a upper bound for the critical quark mass, i.e. $am_c \leq0.00039(1)$ which
corresponds to a upper bound of the critical pion mass of 50 MeV. This is consistent with the results obtained from the calculations using stout fermions~\cite{Endrodi:2007gc} and Wilson fermions 
towards the continuum limit~\cite{Nakamura:2015mea}. On the other hand the fit using the O(2) universality class, in which $m_c=0$, does not accommodate the data at all as seen from the dashed
line in the right plot in Fig.~\ref{fig:Fit3f}.

\section{Summary}

We have studied the chiral phase structure in $N_f$=2+1 and 3 flavor QCD using lattice QCD simulations with various values of quark masses
towards chiral limit using HISQ fermions on $N_\tau=6$ lattices. Our results suggest that the location of the tri-critical point is below that of the physical point,
i.e. $m_s^{tri} < m_s^{phy}$. We found that the transition of $N_f=3$ QCD with bare quark mass $am=0.0009375$ corresponding $m_\pi=80$ MeV in the continuum limit is still
a crossover, and the point where the system starts to have a first order phase transition is estimated to be $am_c \leq 0.00039(1)$ correspoing to $m_\pi^c \lesssim 50 $ MeV.

\end{document}